\documentclass[12pt]{article}
\usepackage{graphicx,amsmath}

\newlength{\dinwidth}
\newlength{\dinmargin}
\def\lapproxeq{\lower .7ex\hbox{$\;\stackrel{\textstyle                                                    
<}{\sim}\;$}}                                                    
\def\gapproxeq{\lower .7ex\hbox{$\;\stackrel{\textstyle                                                    
>}{\sim}\;$}}                                                    
\def\be{\begin{equation}}                                                    
\def\ee{\end{equation}}                                                    
\def\bea{\begin{eqnarray}}                                                    
\def\eea{\end{eqnarray}}                                                    
\def\GeV{\rm GeV}

\def\xg{{\rm in ~the~ largest}~ x_{\gamma} ~{\rm bin}}
\begin{document}                                                    
\titlepage                                                    
\begin{flushright}                                                    
IPPP/09/89  \\
DCPT/09/178 \\                                                    
\today \\                                                    
\end{flushright}                                                    
                                                    
\vspace*{0.5cm}                                                    
                                                    
\begin{center}                                                    
{\Large \bf Factorization breaking in diffractive dijet}\\ 
\vspace*{0.2cm}
{\Large \bf photoproduction at HERA}                                                                                                        
                                                    
\vspace*{1cm}                                                    
A.B. Kaidalov$^{a}$, V.A. Khoze$^{b,c}$, A.D. Martin$^{b}$ and M.G. Ryskin$^{b,c}$  \\                                                    
                                                   
\vspace*{0.5cm}                                                    
$^a$ Institute of Theoretical and Experimental Physics, Moscow, 117259, Russia\\
$^b$ Institute for Particle Physics Phenomenology, University of Durham, Durham, DH1 3LE \\                                                   
$^c$ Petersburg Nuclear Physics Institute, Gatchina, St.~Petersburg, 188300, Russia            
\end{center}                                                    
\vspace*{1cm}                                                    
                                                    
\begin{abstract}
We discuss the factorization breaking observed in diffractive dijet photoproduction by the H1 and ZEUS collaborations at HERA. By considering the effects of rapidity gap survival, hadronisation, migration and NLO contributions, we find that the observed data are compatible with theoretical expectations.
\end{abstract}

\section{Introduction}
As is well known, in QCD the cross section for a `hard' inclusive process {\it factorizes} into universal parton densities and calculable hard  subprocess cross sections. However, for diffractive processes the factorization into universal diffractive parton densities and the known subprocess cross sections may be {\it broken}, since the rapidity gaps associated with the diffractive process can be populated by secondary particles from `soft' rescattering.

Here, we discuss factorization breaking of the diffractive photoproduction of dijets at HERA\footnote{The HERA data for the diffractive production of dijets in deep inelastic scattering are consistent with NLO predictions \cite{H1,JHEP,ZEUSdis}. That is,  within the present uncertainties of the data and theory, no evidence of factorization breaking is observed.}, where each jet has large transverse momentum, $E_T$. That is,  
 events with a large rapidity gap between the proton and the 
hadronic dijet system produced by a photon with virtuality $Q^2 \sim 0$ (see for example \cite{H1,ZEUS}). The domains $x_P<0.03$ and $x_P<0.025$ were selected by the H1 and ZEUS collaborations, respectively. The comparison of these data with theory is not well understood. The situation may be summarised as follows.  In general, there is a tendency for the observed cross sections to be smaller than those predicted \cite{KK,KK2,FS} by NLO QCD. Indeed, for the H1 choice of jet cuts, $E_{T{\rm jet}1}>5$ GeV ($E_{T{\rm jet}2}>4$ GeV), the ratio of data/theory is about $0.5-0.6$ independent of the observed $x_{\gamma}$ \cite{H1}, indicating an overall suppression relative to the QCD prediction.  On the other hand, with higher jet cuts, $E_{T{\rm jet}1}>7.5$ GeV ($E_{T{\rm jet}2}>6.5$ GeV), the data of the ZEUS collaboration \cite{ZEUS} give a data/theory ratio of 0.9 if the diffractive PDFs of H1 fit B \cite{dpdf} are used, compatible with little or no overall suppression\footnote{In a recent ZEUS paper \cite{ZEUSddijets} the data on diffractive dijet production in DIS have been included in a DGLAP analysis to better constrain the diffractive gluon densities. With these diffractive PDFs the dijet data are well described without factorization breaking, but see the comments in the concluding Section.}. Moreover, in the latest H1 analysis \cite{Cerny} the events have been selected using a similar set of cuts to those adopted by ZEUS, with, in fact, identical choices of the $E_T$ and $x_P$ cuts. In this case, the (preliminary) H1 results give, using the H1 `jets' diffractive PDFs \cite{JHEP}\footnote{If H1 fit B diffractive PDFs are used then the ratio for the preliminary H1 data implies an overall suppression of about 0.6 \cite{Cerny}.}, a data/theory ratio consistent with an overall suppression of about 0.8 \cite{Cerny}, which, within the $20-30\%$  experimental uncertainties, is not in contradiction with the findings of ZEUS.

There was an attempt to describe the factorization breaking by soft spectators from the photon interacting inelastically with the proton target and producing secondaries which populate the rapidity gap. For the hadron-like component of the photon wave function this would have produced a suppression by a factor of about 3, corresponding to a gap survival factor $S^2=0.34$ \cite{KKMR}. This idea was widely discussed and used, for example, in the studies of Klasen and Kramer \cite{KK,KK2}. Nevertheless, the absorption of the hadron-like component of the photon cannot explain the suppression of the dijet yield observed $\xg$, close to $x_{\gamma}=1$. When the dijet system carries away almost all of the incoming photon momentum, the `direct $\gamma \to$ dijet' subprocess dominates and we anticipate that $S^2\simeq 1$, since the absorptive cross section of the point-like direct photon is very small.

\section{Components of the photon}
Before continuing the discussion, it is useful to elaborate the different components of the photon, since they are partly a matter of convention. First, we have the `direct' and `resolved' contributions of photon interactions. However, recall that the PDFs which describe the parton distributions of the `resolved' photon contain not only the `hadron-like' component, but
also a so-called `point-like' contribution which originates from the inhomogeneous term in the DGLAP evolution equation for the 
photon \cite{wit}. From a Feynman-graph viewpoint, this `point-like' contribution looks just like the diagrams for the `direct' photon interaction. However, the generally accepted convention is to include the part of this contribution which corresponds to virtualities, $|q^2|$, lower than the factorisation scale $\mu^2_F$,  in the PDFs of the
`resolved' photon, while the part corresponding to larger virtualities, $|q^2|>\mu^2_F$, is included in the NLO matrix element of the `direct' photon interaction. 

Unlike the `hadron-like' component of the 
resolved photon, the probability of an additional soft interaction for the `point-like' component is rather small. It is driven by the size of the $q\bar q$ dipole produced by the initial photon, $\gamma\to q\bar q$.
Thus the absorptive effect 
 for the `point-like' component of the photon PDFs is not so strong; 
we expect
the gap survival factor $S^2$ to be rather close to 1.
 The fact that the `point-like' component of the resolved photon
has a different absorptive cross section to that for the `hadron-like' part, and therefore that the `point-like' component should have a different gap survival factor $S^2$, was discussed first in \cite{KK}. 

Moreover, a detailed study of the photon PDFs, that have been used to make predictions, show that practically in the whole of the observed $x_\gamma$ region selected by the H1 and ZEUS cuts (that is, cuts on $p_{T{\rm jet}}$ and $\eta_{\rm jet}$, which lead to $x_\gamma >0.1-0.2$), the contribution which  originates from the point-like photon dominates (see Fig. 4 of GRV \cite{GRV}). As mentioned above, this component of the photon wave function has no soft spectators and the only absorption comes from the rescattering of the relatively small-size $q\bar{q}$ dipole that is directly produced by the $\gamma \to q\bar{q}$ transition via the inhomogeneous term in the DGLAP evolution of the photon PDFs.  To account for the absorptive effect, this inhomogeneous contribution should be multiplied by the well-known  eikonal
 factor exp$(-\sigma_{\rm abs}/R^2)$, which is of 
analogous form to that introduced in the saturation model of \cite{GBW}. Here, the absorption cross section, $\sigma_{\rm abs}$, depends on the size of the pair, which, in turn, is driven by the running scale $\mu$ in the DGLAP evolution, that is $\sigma_{\rm abs} \propto 1/\mu^2$. In other words, in diffractive processes the point-like photon contribution should be suppressed by $S^2 \sim$ exp$(-a/\mu^2)$ {\it inside} the DGLAP evolution, where the value of the parameter $a$, namely $a \sim 0.6~ \GeV^2$, has been obtained from the known $\sigma_{q\bar{q}-p}$ of the dipole model\footnote{This value, $a\simeq m_\rho^2$, is consistent with the dipole parametrizations of Refs. \cite{DM1,DM2, DM3}.}. Clearly this suppression is only non-negligible near the starting scale of the evolution, $\mu=\mu_0$.

So to summarize:
\begin{itemize} 
\item for the {\it hadron-like component} we expect $S^2\sim 0.34$,
\item for the {\it direct photon contribution} we have $S^2=1$,
\item for the {\it point-like component of the resolved photon}, generated by the inhomogeneous DGLAP term, $S^2$ should be accounted in the DGLAP evolution. That is, it will be of `power-correction' form, $S^2\simeq1-a/\mu^2+...$, which will only be non-negligible at the beginning of the evolution.
\end{itemize}
In the existing photon PDF analyses this power correction is neglected. In the more recent `AFG' photon PDF analysis \cite{AFG}, where the optimum starting scale was $\mu_0^2=0.7 ~\GeV^2$, the corresponding effect is practically negligible ($< 20\%)$. On the other hand, we expect that the use of the `GRV' \cite{GRV} photon PDFs, which have a starting scale of $\mu_0^2=0.3 ~\GeV^2$, will overestimate the point-like part of the resolved photon contribution (by up to about 25$\%$).

\section{Interpretation of HERA data}
We return to the discussion of the measurements of the diffractive photoproduction of dijets at HERA that we outlined in the Introduction.
After all the cuts have been made, the point-like component of the photon dominates (except in the largest $x_\gamma$ bin, where 70$\%$ comes from the direct component\footnote{This can be estimated in various ways -- for example, from the results of Monte Carlos or by inspection of the results in \cite{KK}.}). We therefore expect only weak absorption throughout the observed $x_\gamma$ interval. Indeed, this expectation of only a small amount of factorization breaking for the higher $E_T$ dijet data (with $E_{T{\rm jet}1}>7.5$ GeV  and $E_{T{\rm jet}2}>6.5$ GeV) data is, within the uncertainties, compatible with the observations of ZEUS \cite{ZEUS} and with an overall suppression of about 0.8 seen in the preliminary H1 data \cite{Cerny}\footnote{Recall that we noted that the ZEUS \cite{ZEUS} and preliminary H1 data \cite{Cerny} were obtained using H1 fit B \cite{dpdf} and H1 `dijet' \cite{JHEP} diffractive PDFs respectively.}. Moreover, there is a (weak) tendency of the ratio of data/theory to be larger in the largest $x_\gamma$ bin, though the variation is within the uncertainties of the ratio. 

Most probably, in the largest $x_\gamma$ bin, we observe the interplay of a few different physical effects. First, the GRV photon PDFs \cite{GRV} used in the theoretical calculations, overestimate the effect of the point-like contribution of the resolved photon due to a too small starting scale $\mu^2_0$. Second, as described above, there should be some absorption in the inhomogeneous DGLAP term at the beginning of the evolution. Even using the AFG photon PDFs \cite{AFG}, we expect a bit of absorption. From these two effects, we expect some weak factorization breaking in the region where the {\it resolved} photon contribution dominates. However, in the largest $x_\gamma$ bin, we might anticipate no suppression for the {\it direct} photon component. On the other hand, the hadronization corrections estimated from the LO RAPGAP Monte Carlo \cite{rapgap}, and applied to the NLO  theoretical calculations of Refs. \cite{KK,KK2,FS}, are rather large. In comparison with the value of $x_\gamma$ calculated at the parton level, after hadronization the fraction $x_\gamma$ of the photon's momentum carried by the dijet system becomes smaller. Events from the largest $x_\gamma$ bin migrate to lower $x_\gamma$. Therefore, we expect a negative correction due hadronisation in the largest $x_\gamma$ bin and a positive correction for $x_\gamma \sim 0.7$. Thus, the final effect brings the event rate in the largest $x_\gamma$ bin into line with the other $x_\gamma$ bins, providing, surprisingly, an almost uniform suppression independent of $x_\gamma$ for $x_\gamma > 0.2$.

 Recall that for large photon virtualities, that is, for diffractive high-$E_T$ dijet production in deep inelastic scattering (DIS), the data \cite{H1,JHEP,ZEUSdis} are
consistent, within uncertainties, with the NLO predictions without any factorisation breaking.
Here we deal with a very small-size incoming $q\bar q$ dipole and so the absorptive effects are negligible\footnote{ 
Note that there is no suppression and, correspondingly, no factorisation breaking, in the case of diffractive charm photoproduction \cite{charm} where: (a) the initial heavy-quark  small-size dipole, $c\bar c$, has a small absorptive cross section, and (b) the value of $x_\gamma$ is not fixed -- so hadronisation does not affect the results. Next, recall that the ZEUS data \cite{ZEUS}, obtained with the larger $E_T$
     threshold, are consistent with no, or little, suppression 
in the whole observed range of  $x_\gamma$.}. For this DIS case, it would be interesting to study the role of the hadronisation and migration as a function of $x_\gamma$. Unfortunately, in \cite{H1,JHEP} only cross sections
integrated over $x_\gamma$ are presented, so the effects of migration are not revealed. The diffractive DIS $x_\gamma$ dijet distribution was shown in \cite{ZEUSdis}, see also \cite{H1dijet}.  At large $x_\gamma$ the data agree with the RAPGAP LO Monte Carlo. However, the NLO
calculation leads to a larger cross section than that at LO level. Indeed, in Fig. 6(d) of \cite{ZEUSdis} the NLO prediction in the largest $x_\gamma$ bin exceeds the data by about 20$\%$ or more. This {\it may} be interpreted as an indication in favour of migration  (similar to that in photoproduction) which decreases the observed value of $x_\gamma$.

We also note, from detailed studies made by Klasen and Kramer \cite{KKp}, that at the NLO level, a rather large fraction of the events correspond to three-jet production. For example, the fraction is 
up to 30\%  if the factorisation scale is chosen to be $\mu_F=E_T$
and the cut on transverse momentum of the third jet is taken to be $p_{3T}>1$ GeV.
Of course, in three-jet events, the two highest $E_T$ jets cannot carry
$x_\gamma=1$. On the other hand, the value,  $x_{3\gamma}$,  of momentum fraction carried by the third (lowest $E_T$) jet, cannot be too large. So these three-jet events go mainly to the bins with $x_\gamma=0.6-0.9$.
 Moreover, for the large factorisation scale $\mu_F=E_T$, 
these three-jet events are in an azimuthal `Mercedes'-like configuration.
This is unlike the additional jets generated as parton showers
by RAPGAP (and other) Monte Carlos, which are collinear to a
high $E_T$ or beam jet. Thus, the hadronisation corrections 
calculated based on the LO
 Monte Carlo should be different from the effects associated with 
NLO contribution. In the `Mercedes'-like NLO configuration it is natural to expect stronger migration from the highest $x_\gamma$ bin to smaller values of $x_\gamma$.  

Note that using lower $E_T$ cuts on the jets, we expect both the
hadronisation corrections and the absorptive effect to increase.
As a rule, the non-perturbative
hadronisation correction reduces the value of $E_T$ by some constant amount
$\Delta E_T$. Therefore the effect of
hadronisation is expected
to be stronger for lower $E_T$.
Next, a larger part of  the DGLAP evolution
is affected by the absorption of the $q\bar q$ dipole for lower $E_T$. This is because
the dipole size, $\sim 1/\mu$ with $\mu\sim a$, is not small in a larger part
of the
interval in $\ln\mu^2$ space going from $\mu_0$ up to $E_T$, see eq. (\ref{eq:a1}) and footnote 7 below. 
Indeed, for the lower choice of $E_T$ cuts, the data/theory ratio is observed to be smaller ($\sim 0.5-0.6$) \cite{H1}. However, only a minor part of this suppression is predicted to come from soft rescattering, $S^2$.

For a simple estimate we may use $S^2={\rm exp}(-a/\mu^2)$ with $a=0.6~\GeV^2$ to calculate the expected suppression of the point-like component of the resolved photon. We use the `AFG' photon PDFs \cite{AFG}. Then for the quark- and gluon-partons we find\footnote{Using the `GRV' photon PDFs \cite{GRV} the corresponding predictions are $S^2_q=0.75(0.71),~~S^2_g=0.58(0.53)$.}
\be
S^2_q=0.84(0.81),~~~~~~S^2_g=0.74(0.71)
\label{eq:a1}
\ee
for data samples with $E_{Tjet1}>7.5(5)$ GeV. The quark contribution dominates for large $x_\gamma$, and even for $x_\gamma=0.2$ the gluon contribution is only 20$\%$. Thus, using the value of $S_q^2$, we expect the absorptive suppression to be $S^2 \sim 0.8$, except in the largest $x_\gamma$ bin, where the direct contribution (with $S^2=1$) dominates; the resolved contribution from the quark gives only 30$\%$ so the final prediction is $S^2 \sim 0.94$ in this bin. Then we have to consider hadronisation and the effects of migration. Of course it would be better to implement the NLO `$2\to 3$' matrix element into a Monte Carlo
generator (say, RAPGAP) and to trace the result of hadronisation of the three-jet system explicitly\footnote{In the ideal case, the absorption of the incoming $q\bar q$ dipole in the point-like part of the resolved photon component, that is the factor $S^2=\exp(-a/\mu^2)$, should be included in this Monte Carlo as well.}. At the moment we have no such possibility. However, it is reasonable to assume that due to the non-negligible 
three-jet contribution, the hadronisation correction should be 
about 10\% stronger than that given by the present RAPGAP,
decreasing (by about 10\%, as indicated by the hadronization corrections in the RAPGAP LO Monte Carlo) the NLO prediction for largest $x_\gamma$ bin and enlarging (by about 5\%) the predictions for
 lower $x_\gamma$. This, together with the $S^2$ suppression of the point-like component of the resolved photon, can satisfactorily describe the present experimental data.

\section{Conclusions}

If we were to neglect all final state interactions, and have no hadronisation corrections or migration effects, we would have exact factorisation in the DIS diffractive production of dijets, but some factorisation breaking in diffractive photo-produced dijets where the hadron-like component of the resolved photon suffers a small gap survival probability, for example, $S^2 \simeq 0.34$ \cite{KKMR}. However in the kinematic region of the HERA dijet data, $x_\gamma \gapproxeq 0.1$, the hadron-like contribution is very small. Here the point-like part of the resolved photon dominates, for which $S^2$ is close to one, see (\ref{eq:a1}).

We have argued that the factorisation breaking observed in the diffractive photoproduction of dijets is most probably caused by hadronisation corrections and migration effects in the final state. These effects should be similar in both  DIS- and photo-produced diffractive dijets. For this reason it is dangerous to include DIS diffractive dijet data in diffractive PDF (DPDF) analyses. The analyses assume exact factorisation and therefore tune the DPDFs to partly compensate for the effect of the interactions in the final state. Note that after including DIS diffractive dijet data in the DPDF analysis, ZEUS \cite{ZEUSddijets} do not observe factorisation violation in diffractive photoproduced dijets. This observation may be considered as support for our argument. That is, the factorisation breaking, observed when diffractive dijet photoproduction data are described using DPDFs obtained by analysing pure inclusive diffractive DIS data (such as H1 fit B \cite{dpdf} or MRW \cite{MRW}), is mainly of hadronisation/migration origin.

In summary, the hadron-like component of the resolved photon, which is suppressed by a factor $S^2 \simeq 0.34$ \cite{KKMR}, only starts to be important for small $x_\gamma$. Indeed, to feel the hadron-like component one needs to observe dijets far in rapidity from the photon, corresponding to $x_\gamma<0.1$. This region was difficult to access at HERA\footnote{This statement is not strictly true. For example, in Ref. \cite{H1old}, the H1 collaboration presented results down to $x_{\gamma}=0.05$ for dijets with jets of $E_{T\rm jet}>4$ GeV and $|\eta_{\rm jet}|<2.5$.}. The point-like component of the resolved photon, which is calculable perturbatively, is the dominant one for $x_\gamma>0.1$, and has a small suppression. For this component, the spectator partons have relatively large transverse momenta and can be seen as a third jet. Finally, after including the direct component and taking into account the effects of hadronisation and migration, we find that our expectations are consistent with the observed data for diffractively photoproduced dijets.
  
\section*{Acknowledgements}
We thank Alice Valkarova for rekindling our interest in this problem. We are grateful to her and to Paul Newman for valuable discussions of experimental issues. We especially thank Michael Klasen and Gustav Kramer for providing NLO distributions for diffractive dijet photoproduction. MGR would like to thank the IPPP at the University of Durham for hospitality. This work was supported by the grant RFBR
07-02-00023, by the Federal Programme of the Russian State RSGSS-3628.2008.2.


\thebibliography{} 


\bibitem{H1}
H1 collaboration, A.~Aktas {\it et al.},
in
  Eur.\ Phys.\ J.\  {\bf C51}, 549 (2007).
\bibitem{JHEP} H1 collaboration, A. Atkas {\it et al.}, JHEP {\bf 0710}:042 (2007).
\bibitem{ZEUSdis} ZEUS collaboration,  S.~Chekanov {\it et al.},
  Eur.\ Phys.\ J.\  {\bf C52}, 813 (2007).

\bibitem{ZEUS} ZEUS collaboration,  S.~Chekanov {\it et al.},
 Eur.\ Phys.\ J.\  {\bf C55}, 177 (2008).



\bibitem{KK} M. Klasen and G. Kramer, Eur. Phys. J. {\bf C38}, 93 (2004); Proc. of 12th International Workshop on Deep Inelastic Scattering, \v{S}trbsk\'{e} Pleso, (DIS 2004) p.492,  hep-ph/0401202.
\bibitem{KK2}  M. Klasen and G. Kramer, Proc. of HERA and the LHC workshop series (2006-2008), p. 448, DESY-PROC-2009-02;  Mod. Phys. Lett. {\bf A23}, 1885 (2008) [arXiv:0806.2269, hep-ph] and references therein.

\bibitem{FS} S. Frixione and G. Ridolfi, Nucl. Phys. {\bf B507}, 315 (1997).

\bibitem{dpdf} H1 collaboration, A. Aktas {\it et al.}, Eur. Phys. J. {\bf C48}, 749 (2006).

\bibitem{ZEUSddijets} ZEUS collaboration, S. Chekanov {\it et al.}, arXiv:0911.4119

\bibitem{Cerny} 
 K.~Cerny,
 16th International Workshop on Deep Inelastic
Scattering
and Related Subjects (DIS 2008), London, England, April 2008;\\
A. Valkarova, to be published in the Proc.
of Hadron Structure 09.

\bibitem{KKMR} A.B. Kaidalov, V.A. Khoze, A.D. Martin and M.G. Ryskin, Phys. Lett. {\bf B567}, 61 (2003).
\bibitem{wit} 
 E. Witten,  Nucl. Phys. {\bf B120}, 189 (1977).

\bibitem{GRV} M. Gluck, E. Reya and A. Vogt, Phys. Rev. {\bf D46}, 1973 (1992).

\bibitem{GBW} 
 K.J. Golec-Biernat and M. Wusthoff,  Phys. Rev. {\bf D59}, 014017 (1999). 


\bibitem{DM1}  L. Motyka, K.J. Golec-Biernat and G. Watt, Proc. of HERA and the LHC workshop series (2006-2008), p. 471, DESY-PROC-2009-02
[arXiv:0809.4191, hep-ph]. 

\bibitem{DM2} J. Bartels, K.J. Golec-Biernat and H. Kowalski, Phys. Rev. {\bf D66}, 014001 (2002). 

\bibitem{DM3} J. Nemchik, N.N. Nikolaev, E. Predazzi and B.G. Zakharov, Phys. Lett. {\bf B374}, 199 (1996).

\bibitem{AFG} P. Aurenche, M. Fontannaz and J.Ph. Guillet, Eur. Phys. J. {\bf C44}, 395 (2005).
\bibitem{rapgap} 
 H. Jung, DESY-93-182, Comput. Phys. Commun. {\bf 86}, 147 (1995).


\bibitem{charm} H1 collaboration, A. Atkas {\it et al.}, Eur. Phys. J. {\bf C50}, 1 (2007),\\
ZEUS collaboration, S. Chekanov {\it et al.}, Nucl. Phys. {\bf B672}, 3 (2003);
 Eur. Phys. J. {\bf C51}, 301 (2007).
 
\bibitem{H1dijet} H1 collaboration, C. Adloff {\it et al.}, Eur. Phys. J. {\bf C20}, 29 (2001).

\bibitem{KKp} M. Klasen and G. Kramer, private communication.

\bibitem{H1old} H1 collaboration, C. Adloff {\it et al.}, Phys. Lett. {\bf B483}, 36 (2000).

\bibitem{MRW} A.D. Martin, M.G. Ryskin and G. Watt, Phys. Lett. {\bf B644}, 131 (2007).
\end{document}